\documentclass{article}
\usepackage{latexsym}
\begin{document}

\title{The Profound Spinor Combination of Space-time-matter}
\author{Lun Biao Xu \\ \it Physics Department, Zhejiang University, Hangzhou, 310027, P.R.China\\ \it Address: Qiu-Shi-Cun 34-502, Hangzhou 310013, P.R.China}
\maketitle

\begin{abstract}
Following the famous Dirac equation, in which space, time and matter are all connected with spinor, this paper uses the combination of these spinors to express the state of quantum field in a new style - the global state. Thus, the state, state equation and action function of quantum fields are all expressed in an integrated system. Based upon this, it seems that the origin of the principle of least action, the intrinsic connection between classical mechanics and quantum theory can be fully exposed. 
 
\it Key words: Space-time-matter, Spinor combination, Classical and quantum field equation

\end{abstract}

\vspace{8pt}
\noindent {\Large \bf 1 Introduction}
\vspace{4pt}

On the concept of space-time, two levels of understanding have been reached in the 20th century. The first is the universality of space-time, which means that space and time are not absolutely independent but are relatively connected with each other. The second is that there are two representations for demonstrating the fashion of every matter: the space-time representation and momentum-energy representation, which are wave-particle dual representations. In the universe, the space-time-matter originally has its twin, the canonical conjugate momentum-energy-mass. This can be called trinity with dual aspects. 

In fact, by examining physics in the 20th century carefully, a third breakthrough in the developing viewpoint of space-time and matter movement can be discerned. The state of microscopic particles has been found relating with spinor since 1928 when Dirac represented his relativistic quantum theory of electrons [2, 8, 15, 5]. Although this concept, space-time is related to spinor, has been wildly used in quantum fields theory, it seems that it still can be further developed in a more effective way. Meanwhile, there are still many problems remained unresolved in physics in last century. For example, how to express the state of any field clearly and simply in terms of space-time-matter or momentum-energy-mass.

In this paper, starting from Dirac spinor, this third level of understanding will be thoroughly investigated. Using ``the space-time-matter spinor pattern'', all fields, including their states and movements are expressed concisely and unitedly under a grand system. The whole investigation demonstrates the perfect unification of relativity, quantum theory and gauge field theory, as well as the intrinsic connection of classical mechanism and quantum theory. 

\vspace{8pt}
\noindent {\Large \bf 2 The essence of Dirac equation; the states and motion equations for particle-field in Spinor pattern}
\vspace{4pt}

Dirac presented the free electron's Schr\"{o}dinger equation $(\hat{H}-E)\psi=0$ in the form $(i\partial_t-\alpha \cdot p- \beta m)\psi=0$, where $\alpha =\sigma ^1 \otimes \sigma, \beta = \sigma ^3 \otimes I$, $\sigma^1, \sigma^2, \sigma^3$ are Pauli matrixes.(In this paper, $c=\hbar=e=1$, and for simplicity, some constants such as $2\pi$ are omitted in Fourier transformation, in phase terms or etc.). 

The great advance of this form is that the three amount $m, \epsilon, p$ are put in different spinor matrix. Then $m, \epsilon, p$ construct a coordinate in spinor space. Applying the relativistic relation $(H-E)(H+E)=H^2-E^2=\epsilon^2-p^2-m^2=0$, it can be learned that the eigenstate $\psi$ of $(H-E)$ can be obtained from $(H+E)$ matrix [8, 5], i.e. the four 4-components spin states are the four columns of this $4\times 4$ matrix. In another word, $(H+E)$ is the combination of four eigenstates of electron, or global state. By recasting Dirac equation, two important hints can be got. First, the global momentum eigenstate of free electron can be expressed in a special spinor combination, or spinor pattern, which is constructed when $m, \epsilon, p$ are put at each spinor coordinate and added together. Second, in momentum representation, the state equation is $(H-E)(H+E)=0$, or $(H+E)(H-E)=0$. This means the eigenstate of operator $(H-E)$ is $(H+E)$ , and the eigenstate of operator $(H+E)$ is $(H-E)$. This form not only expresses the existence of $\pm E$ energy electron, but also expresses the existence picture of plus-minus particles in vacuum, which can be named as Hun-tun equation.

But Dirac equation can be further improved. In Dirac equation, one of $m, \epsilon, p$ is put in $I$ (unite matrix) place. This makes the spinor coordinates of $m, \epsilon, p$ unequal at Clifford algebra [17, 10]. Referring to Kaufman's spinor system [12] (he presented an expression system of spinor group to improve the Ising model in 1949), a suitable transformation from Dirac $\alpha - \beta - I$ system can be made to the STM (space-time-matter) spinor system: 

$\gamma ^m=i \sigma ^3 \otimes I$ 

$\gamma ^\epsilon = -\sigma ^1 \otimes I$

$\gamma ^i = i \sigma ^2 \otimes \sigma ^i \hspace{3em} (i=1,2,3)$ \hfill (1)

Then the spin $S=\frac{1}{2}$ particle-field was expressed as a spinor pattern state $P(p)$:

$P(p)\equiv P\left( \begin{array}{cc} \gamma ^m & m \\ \gamma ^\epsilon & \epsilon \\ \gamma ^i & p_i \end{array} \right) \equiv (\gamma ^m m+ \gamma ^\epsilon \epsilon + \gamma ^i p_i)$

$\qquad \equiv (\gamma ^m m + \gamma ^\epsilon \epsilon + \gamma ^1 p_1 + \gamma ^2 p_2 +\gamma ^3 p_3)$ \hfill (2)

Corresponding to Dirac equation $(H\pm E)(H\mp E)=0$, the state equations are as follows:

$P(p)\bar{P}(p)=0$ \qquad and \qquad $\bar{P}(p)P(p)=0$ 
\hfill(3)

where $\bar{P}(p)=-P(p)$ 

Above are the momentum eigenstate and state equation (or motion equation) for the free $S=\frac{1}{2}$ particle-field in the momentum representation. Referring to quantum mechanics, the expression of momentum eigenstate in coordinate representation is obtained [9]:

$\left\langle Q(x)|P(p) \right\rangle = P(p)e^{iP(p)Q(x)}$ \hfill(4)

where $Q(x)$ is the $s, t, x$ spinor pattern, canonical conjugate with $P(p)$.

$Q(x) \equiv Q \left( \begin{array}{cc} \gamma ^m & s \\ \gamma ^\epsilon & t \\ r^i & x_i \end{array} \right)\equiv Q(\gamma ^ms + \gamma ^\epsilon t + r^i  x_i)$

$\qquad \equiv (\gamma ^ms + \gamma ^\epsilon t + r^1x_1+r^2x_2+r^3x_3)$ \hfill(5)

It can be noticed that the phase is written in spinor, but not scalar as in traditional form. This can be confirmed since angular $L=p\times x$ term is expressed from the closed terms of $P(p)Q(x)$, and moreover, the spinor phase term includes deep and plentiful quantum information. 

\vspace{8pt}
\noindent {\Large \bf 3 The Maxwell field in spinor pattern}
\vspace{4pt}

Maxwell field (EM field) is a kind of gauge field, which is commonly expressed in four dimension tensor: $\partial _\mu F^{\mu \nu}=0$ or $\partial _\mu F^{\mu \nu}=j^\nu$, for free and source field respectively, where $F^{\mu \nu}=\partial ^\mu A_\nu - \partial ^\nu A_\mu$ [1, 16, 7].

If the particle field $P(p)$ in equation (3) is looked as the excitation of first order spinning in vacuum, the equation of $F^{\mu\nu}$ shows that gauge field is the existence of second order spinning in vacuum. Expressing this in the STM spinor pattern, the state of Maxwell gauge field is:

$P(A)=P\left( \begin{array}{cc} \gamma ^\epsilon & \varphi (x) \\ \gamma ^i & A_i(x) \end{array} \right)$ \hfill(6)

without the $\gamma ^m$ term. The field equation is:

$P(\hat{p})P(\hat{p})\bar{P}(A)=0$ \hfill(7)

where $P(\hat{p})\equiv P\left( \begin{array}{cc}\gamma ^\epsilon & i\partial _t \\ \gamma ^i & -i\partial _i \end{array}\right)\equiv \left[\gamma ^\epsilon (i\partial _t)+\gamma ^i(-i\partial _i)\right]$

$\qquad \equiv \left[\gamma ^\epsilon (i\partial _t)+\gamma ^1(-i\partial _1)+\gamma ^2(-i\partial _2)+\gamma ^3(-i\partial _3)\right]$ \hfill(8)

Precisely, 

$P(\hat{p})P(\hat{p})\bar{P}(A(x))=P\left( \begin{array}{cc}\gamma ^\epsilon & i\partial _t \\ \gamma ^i & -i\partial _i \end{array}\right)P\left( \begin{array}{cc}\gamma ^\epsilon & i\partial _t \\ \gamma ^i & -i\partial _i \end{array}\right)\bar{P}\left( \begin{array}{cc}\gamma ^\epsilon & \varphi (x) \\ \gamma ^i & A_i(x) \end{array}\right)$

$\qquad =P\left(\begin{array}{cc}\gamma ^\epsilon & i\partial _t \\ \gamma ^i & -i\partial _i \end{array}\right)[-\gamma ^\epsilon \gamma ^\epsilon(i\partial _t\varphi)-\gamma ^i\gamma ^i(-i\partial _iA_i)-\gamma ^i\gamma ^j(-i\partial _iA_j+i\partial _jA_i)$

$\qquad\qquad -\gamma ^\epsilon\gamma^i(i\partial_tA_i)-\gamma ^i\gamma ^\epsilon(-i\partial_i\varphi)]$

$\qquad =P\left( \begin{array}{cc}\gamma ^\epsilon & i\partial _t \\ \gamma ^i & -i\partial _i \end{array}\right)\left[-I\otimes I(iG) +I\otimes\sigma\cdot(B) + \sigma^3\otimes\sigma\cdot(iE)\right]$

$\qquad = - \sigma^1\otimes I(\partial_tG)-i\sigma^1\otimes\sigma\cdot(\partial_tB)-i\sigma^2\otimes\sigma\cdot(\partial_tE)-i\sigma^2\otimes\sigma\cdot(\nabla G)$

$\qquad \qquad +\sigma^2\otimes I(\nabla\cdot B)-\sigma^1\otimes I(\nabla\cdot E)+i\sigma^2\otimes\sigma\cdot(\nabla \times B)-i\sigma^1\otimes\sigma\cdot(\nabla \times E)$

$\qquad =\left[21\right]+\left[12\right]=0$ \hfill(9)

where $\left[21\right]$ and $\left[12\right]$ are the left-down and right-up quarter of the whole $4\times4$ matrix respectively,

$\left[12\right]=-I(\partial_tG)-\sigma\cdot(\nabla G)-I\cdot(\nabla\cdot E)-\sigma\cdot(\partial_tE-\nabla\times B)-iI(\nabla\cdot B)$

$\qquad \qquad -i\sigma\cdot(\partial_tB+\nabla\times E)=0$

$\left[21\right]=-I(\partial_tG)+\sigma\cdot(\nabla G)-I\cdot(\nabla\cdot E)+\sigma\cdot(\partial_tE-\nabla\times B)+iI(\nabla\cdot B)$

$\qquad \qquad -i\sigma\cdot(\partial_tB+\nabla\times E)=0$ \hfill(10)

where $G\equiv\partial_t\varphi+\nabla\cdot A,\qquad E\equiv-\partial_tA-\nabla\varphi,\qquad B\equiv\nabla\times A$

$\left[21\right]$ and $\left[12\right]$ are both constructed by real and imaginary parts, which should be equal to zero respectively. For imaginary part, $\partial_tB+\nabla\times E=0$, because $\nabla\cdot B=\nabla\cdot(\nabla\times A)=0$. This is just Faraday's law. For real part, 

$I(\partial_tG)\pm\sigma\cdot(\nabla G)+I(\nabla\cdot E)\pm\sigma\cdot(\partial_tE-\nabla\times B)=0$, \quad so,

$\left\{\begin{array}{l}\nabla\cdot E=-\partial_tG\equiv\rho \\-\partial_tE+\nabla\times B= \nabla G\equiv j \end{array}\right. $\hfill(11)

So, equation (7) is just Maxwell EM field equation, and $G=0$ or $G\neq0$ is the case of free or source field respectively.

Traditionally, $G$ is always defined as $0$, so the Maxwell equation was expressed in two equations, free and source field respectively. But now, as the gauge field is confirmed to be the presentation of second order spinning in vacuum, its Hun-tun equation is just $P(p)P(p)\bar{P}(A)=0$. It includes both free and source cases, depending on whether $G$ is zero or not. Actually, connecting $\rho$ and $j$ with $(-\partial_tG)$ and $(\nabla G)$, the relationship of $\rho, j$ and $\varphi, A$ is clearer. That is, when there is a divergence $G$ in gauge field, there will be $\rho$ and $j$. Consequently, $\varphi, A$ are all D'alembert's amounts. (When the amount $f$ satisfies $\Box^2f=0$, it is named ``D'Alembert's''.) This fact coincides with the expanding of $\varphi$ and $A$ in harmonic waves.

The meaning of gauge transformation is clearer in the form of spinor pattern equation (7). That is to add a four dimension gradient of arbitrary scalar field $\chi (x)$ on field $\varphi$, $A$, providing $\chi (x)$ is D'Alembert's, the field strength $E$, $B$ and field divergence $G$ are invariant:

$P(\hat{P})P(A)\rightarrow P(\hat{P})P(A')=P(\hat{P})P(A+\nabla\chi)=P(\hat{P})\left[P(A)+P(\nabla\chi)\right]$

\qquad $=P(\hat{P})P(A)+P(\hat{P})P(\nabla\chi)=P(\hat{P})P(A)+P(\hat{P})P(\hat{P})iI\chi$

\qquad $=P(\hat{P})P(A)$\hfill(12)

Consequently, it expounds why the `source' of field $P(j)$ always has $P(\hat{P})P(j)=0$ (current conservation). This is because $P(j)=P(\hat{P})iIG$, and $G$ is D'Alembert's. 

\vspace{8pt}
\noindent {\Large \bf 4 The electron in EM field: an example of interaction field}
\vspace{4pt}

An electron in EM field is an example of interaction field, i.e., the interaction of matter field with gauge field. Now the state and motion equation can be written in spinor pattern.

As well known, the momentum of particle affected by a gauge field should be $\Pi=p-A(x)$, where $\Pi$ is named as mechanical momentum, and $p$ is canonical momentum. From this, the spinor pattern-state of interaction field should be: 

$P(p-A(x))$\hfill(13-1)

and its motion equation in spinor pattern form is: 

$P(p-A(x))\bar{P}(p-A(x))=0$ \hfill(13-2)

where $\gamma^m, \gamma^\epsilon, \gamma^i$ are taken as (1).

This motion equation can be looked as the Hun-tun presentation of first order spinning in the second order space. (In a rough analogy, the Hun-tun presentation of particle happened on a spinning dish.) Then the four-dimension space is curved. Using Einstein's terminology of metric, $g_{\mu\nu}$ is related with $x$-coordinates now, $g_{\mu\nu}(x)\neq\eta_{\mu\nu}$. So the equation of interaction field is general relativistic equation [18, 7].

The expression of equation $P(p-A(x))\bar{P}(p-A(x))=0$ in coordinate representation is a transformation of the operator and state respectively as:

$P(p-A(x))\rightarrow P(-i\nabla-A(x))\equiv P(\hat{p}-A(x))$

$\bar{P}(p-A(x))\rightarrow \left\langle Q(x-B(p))|\bar{P}(p-A(x))\right\rangle$

\qquad $=\bar{P}(p-A(x))e^{i\bar{P}(p-A(x))Q(x-B(p))}$ \hfill(14)

where $x-B(p)\equiv X$ is mechanical coordinate, which conjugated with $\Pi=p-A(x)$; and $B(p)$ is a coordinate value, which conjugated with momentum value $A(x)$. 

Then the motion equation is:

$P(\hat{p}-A(x))\left[\bar{P}(p-A(x))e^{i\bar{P}(p-A(x))Q(x-B(p))}\right]=0$ \hfill(15)

The function in the above bracket is just the wave function of the Dirac electron in EM field [5, 9]. So, with spinor pattern, it is very easy to write the interaction field's equation and its solution.

Moreover, it should be noticed that the coordinate system for every experiment observer who is not affected by gauge field $A(x)$ is only $Q(x)$. So the expression of momentum eigenstate $P(p-A(x))$ in the view-space of experiment observer $Q(x)$ should be 

$\left\langle Q(x)|P(p-A(x))\right\rangle=P(p-A(x))e^{iP(p-A(x))Q(x)}$ \hfill(16)

Essentially, it is the expression of momentum eigenstate $P(p-A(x))$ in mechanical coordinate representation $Q(x)$.

\vspace{8pt}
\noindent {\Large \bf 5 The invariance of state $P(p)$, $P(p-A(x))$ in view-space $Q(x)$}
\vspace{4pt}

$P(p)$ or $P(p-A(x))$ are both momentum eigenstate, whose eigen momentum values are unvaried at any $Q(x)$ point. That is:

$\partial_{Q(x)} P(p)=0$ \hfill(17-1)

$\partial_{Q(x)}P(p-A(x))=0$ \hfill(17-2)

where $\partial_{Q(x)}=P(\hat{p})$. So, for matter field state, it is:

$P\left(\begin{array}{cc} \gamma^m & -i\partial_s \\ \gamma^\epsilon & i\partial_t \\\gamma^i & -i\partial_i \end{array} \right) P\left(\begin{array}{cc}\gamma^m & m \\ \gamma^\epsilon & \epsilon \\ \gamma^i & x_i \end{array} \right)=0$

In the example of $n=2$ spinor pattern, the (17-1) equation is:

$P(\hat{p})P(p)=iI\otimes I(-\partial_sm+\partial_t\epsilon-\nabla\cdot p)-I\otimes\sigma\cdot(\nabla\times P)$

$\qquad -i\sigma^2\otimes I(\partial_s\epsilon+\partial_tm)+\sigma^1\otimes\sigma\cdot(\partial_sp-\nabla m)+i\sigma^3\otimes\sigma\cdot(\partial_tp+\nabla\epsilon)$

$=\left[12\right]+\left[21\right]+\left[11\right]+\left[22\right]=0$ \hfill(18)

where each of $\left[12\right]$, $\left[21\right]$, $\left[11\right]$ and $\left[22\right]$ is a quarter of whole $4\times 4$ matrix, and should be zero, i.e.,

$\left\{\begin{array}{l}i\left[I(-\partial_sm+\partial_t\epsilon-\nabla\cdot p)\pm\sigma\cdot(\partial_tp+\nabla\epsilon)-\sigma\cdot(\nabla\times p)\right]=0 \\ 
\pm I(\partial_s\epsilon+\partial_tm)+\sigma\cdot(\partial_sp+\nabla m)=0 \end{array}\right.$ \hfill(19)

The two possibilities $\pm$ appear in equation, according to spin conditions $\uparrow$ and $\downarrow$. 

These equations show the relationship of space-correlative values (differentiation, gradient, divergence, rotation, etc.) of momentum amount $m$, $\epsilon$ and $p$. They are in fact the classical motion equations of solitary particle-field. When every bracket equals to zero,

$\left\{\begin{array}{l} -\partial_sm+\partial_t\epsilon-\nabla\cdot p=0 \\ \partial_tp+\nabla\epsilon=0 \\ \nabla\times p=0 \\ \partial_s\epsilon+\partial_tm=0 \\ \partial_sp-\nabla m=0 \end{array}\right.$ \hfill(20)

Obviously, these equations include Newton mechanics and the effects such as mass-lose $(\partial_sm)$ terms to the particle-field.

The equation (17-2) is:

$P(\hat{p})P(p-A(x))=0$ \quad or $P\left(\begin{array}{cc}\gamma^m & -i\partial_s \\ \gamma^\epsilon & i\partial_t \\\gamma^i & -i\partial_i \end{array} \right) P\left(\begin{array}{cc}\gamma^m & m-0 \\ \gamma^\epsilon & \epsilon-\varphi(x) \\ \gamma^i & p_i-A_i(x) \end{array} \right)=0$

For simplicity, disregarding terms of $\gamma^m$, and take $n=2$ for example:

$0=P(\hat{p})P(p-A(x))=I\otimes I\left[i\partial_t(\epsilon-\varphi)\right]+\gamma^i\gamma^i\left[-i\partial_i(p-A)_i\right]$

$\qquad\qquad +\gamma^\epsilon\gamma^i\left[i\partial_t(p-A)_i+i\partial_i(\epsilon-\varphi)\right]+\gamma^i\gamma^j\left[-i\nabla\times(p-A)\right]_k$

$\qquad =I\otimes I\left[i\partial_t(\epsilon-\varphi)+i\nabla\cdot(p-A)\right]+\gamma^\epsilon\gamma^i\left[i\partial_t(p-A)_i+i\nabla_i(\epsilon-\varphi)\right]$

$\qquad \qquad +\gamma^i\gamma^j\left[-i\nabla\times(p-A)\right]_k$ \hfill(21)

There are four results, show as follows:

1) $\quad \partial_t\epsilon+\nabla\cdot p=\partial_t\varphi+\nabla\cdot A$

$\qquad \partial_tp+\nabla\epsilon=\partial_tA+\nabla\varphi$

$\qquad \nabla\times (p-A)=0$ \hfill(22)

This expresses that gauge field $\varphi$, $A$ and particle field $\epsilon$, $p$ are mutual ditch-spring, which is a figurative picture of interaction field. Also, it expresses that mechanical momentum $\Pi=p-A$ is rotation-less, as in the previous case of solitary particle-field. 

2) The equations show that the terms of divergence and gradient for particle-field $\epsilon$, $p$ should be discussed. Consequently, in the interaction field, $p$ and $x$ for the particle-field are interwoven. This interweaving is defined in the level at $\partial_xp$.

3) Retaining the term of $\gamma^m(-i\partial_s)$ in $P(p)$, from the equations with $\partial_sm$, $\partial_s(\epsilon-\varphi)$, $\partial_s(p-A)$, the effects of mass-loss or energy-loss on the motion can be discussed.

4) Reducing the variance of $Q(x)$ due to a mono-parameter $t$ in continuous variance, $\Delta Q(x)=Q(\Delta x)=Q(\frac{\Delta x}{\Delta t})\Delta t=\dot{Q}(x)\Delta t$. Consequently, $\partial_{Q(x)}P(p-A(x))=0$ can be written as:

$\frac{\partial P(p-A(x))}{\partial Q(x)}\cdot\frac{\partial Q(x)}{\partial t}=0$

i.e., $\left[P(\hat{p})P(p-A(x))\right]\dot{Q}(x)=0$ \hfill(23)

In fact, this is just the motion equation of classical particle in gauge field. Verifying the case of $n=2$, the equation $\left[P(\hat{p})P(p-A(x))\right]\dot{Q}(x)=0$ is exactly the motion equation of electron in EM field. In detail,

$\left[P(\hat{p})P(p-A(x))\right]\dot{Q}(x)=\left[P(\hat{p})P(p-A(x))\right]Q\left(\begin{array}{cc}\gamma^\epsilon & {\Delta t}/{\Delta t} \\ \gamma^l & {\Delta x_l}/{\Delta t}\end{array}\right)$

$\qquad =\left[P(\hat{p})P(p-A(x))\right](\gamma^\epsilon 1+\gamma^ lv_l)=\left[12\right]+\left[21\right]=0$ \hfill(24)

where $\left[12\right]$ and $\left[21\right]$ are quarters of whole $4\times4$ matrix, which are both zero. Then, there are four equations correspondingly:

Part of $I$: 

\quad $\left[\nabla\times(p-A)\right]\cdot v=0$

$\quad i\left\{\left[-\partial_t(\epsilon-\varphi)-\nabla\cdot(p-A)\right]+\left[\partial_t(p-A)+\nabla(\epsilon-\varphi)\right]\cdot v\right\}=0$

Part of $\sigma$:

\quad $\nabla\times(p-A)-\left[\partial_t(p-A)+\nabla(\epsilon-\varphi)\right]\times v=0$ 

$\quad i\left\{\left[\partial_t(\epsilon-\varphi)+\nabla\cdot(p-A)\right]v\pm\left[\partial_t(p-A)+\nabla(\epsilon-\varphi)+\nabla\times(p-A)\times v\right]\right\}$

$\qquad =0$  \hfill(25)

These four equations explain the particle motion at interaction. Part of the last equation is:

$\partial_t(p-A)+\nabla(\epsilon-\varphi)+\left[\nabla\times(p-A)\right]\times v=0$ \hfill(26)

With some deduction, the familiar motion equation of a charged particle in EM field appears [11, 15, 16]:

$\frac{d}{dt}p=-E-v\times B-\nabla\epsilon+v\cdot\nabla p$, 

At non-relativistic approximation, it becomes the familiar form: 

$\frac{d}{dt}p=-E-v\times B$. \hfill(27)

From the above discussion, the accurate motion equation of classical mechanics is amazingly obtained from a deduction of a complete quantum canonical conjugate relation (17). It is really a profound intrinsic connection. Further in this paper, how the accurate motions of classical mechanics reduces to the quantum results, which is again the expression of this connection, will be demonstrated.

\vspace{8pt}
\noindent {\Large \bf 6 The states and motion equation of any spin fields}
\vspace{4pt}

In previous sections, the electron and Maxwell field are discussed as examples of mater field and gauge field. If broadening the order $n$ of spinor pattern, the same form of state and motion equation can be used at any suitable $S$-spin fields, $P(p)$, $P(A)$, and $P(p-A)$, providing that for matter field, $s=\frac{n-1}{2}$; for gauge field, $s=\frac{n}{2}$. For interaction field, $n$ can be matched with plus $\otimes I$ at the end to equal the order of $P(p)$ and $P(A)$. 

The $n$-order STM spinor system for arbitrary $S$ field is expressed as the following:

$\begin{array}{cc} \gamma^m & i\sigma^3\otimes I\otimes I\otimes\cdots\cdots\otimes I \\ \gamma^\epsilon & -\sigma^1\otimes I\otimes I\otimes\cdots\cdots\otimes I \\\gamma^i & i\sigma^2\otimes\sigma^3\otimes I\otimes\cdots\cdots\otimes I \\ & i\sigma^2\otimes\sigma^1\otimes I\otimes\cdots\cdots\otimes I \\ & i\sigma^2\otimes\sigma^2\otimes \sigma^3\otimes\cdots\cdots\otimes I \\& i\sigma^2\otimes\sigma^2\otimes \sigma^1\otimes\cdots\cdots\otimes I \\& \vdots \\ & i\sigma^2\otimes\sigma^2\otimes \sigma^2\otimes\cdots\cdots\otimes \sigma^3 \\& i\sigma^2\otimes\sigma^2\otimes \sigma^2\otimes\cdots\cdots\otimes \sigma^1\\ & \underbrace{ i\sigma^2\otimes\sigma^2\otimes \sigma^2\otimes\cdots\cdots\otimes \sigma^2} _{n \quad  layers}\end{array}$ \hfill(28)

\vspace{8pt}
\noindent {\Large \bf 7 Phase $\Omega$, action S, and the origin of the principle of least action}
\vspace{4pt}

As discussed previously, the state $P(p-A(x))$ expresses in view representation is:

$\left\langle Q(x)| P(p-A(x))\right\rangle=P(p-A(x))e^{iP(p-A(x))Q(x)}$

where the phase is $\Omega=P(p-A(x))Q(x)$ \hfill(29)

So the phase difference at two points in space-time is:

$S=\Delta \Omega\equiv\int^{Q(2)}_{Q(1)}P(p-A(x))dQ(x)$ \hfill(30)

Actually, this $S$ is just the classical action function, which is not only confirmed by Dirac's opinion and Feynman's practice, but also confirmed by transforming the form of traditional action function $S$ in Lagrange-Hamilton mechanics [6, 14, 3]: 

$S=\int^2_1L(q_i, \dot{q}_i, t)dt=\int^2_1(\sum_{i}^{n}p_i\dot{q}_i-H)dt=\int^2_1(\sum^n_ip_i\dot{q}_i-H\dot{t})dt$

$\qquad =\int^2_1(\sum^{n+1}_i p_i\dot{q}_i)dt=\int^2_1\sum^{n+1}_ip_idq_i$

$\qquad \Rightarrow\int^2_1P(p-A(x))dQ(x)$ \hfill(31)

So the action S can be easily and precisely expressed with spinor pattern $P(p-A)$ and $Q(x)$.

Calculating $S$ with path-integration, the following equation can be obtained:

$S=\int^2_1P(p-A(x))dQ(x)=\left[P(p-A(x))Q(x)\right]^2_1-\int^2_1Q(x)dP(p-A(x))$

$\qquad =\left[P(p-A(x))Q(x)\right]^2_1-\int^2_1Q(x)\left[\frac{\partial P(p-A(x))}{\partial Q(x)}\cdot\frac{\partial Q(x)}{\partial t}\right]dt$

$\qquad =\left[P(p-A(x))Q(x)\right]^2_1$ \hfill(32)

The second integration term equals to zero because $\partial_{Q(x)}P(p-A(x))=0$. 

And $\left[\frac{\partial}{\partial Q(x)}P(p-A(x))\right]\dot{Q}(x)=0$ is just the motion equation of classical particle. This shows the  action is exactly the value-difference of $P(p-A(x))Q(x)$ between points $1\rightarrow 2$ on classical path, or the phase changing on the geodesic line. For instance, the equation $\partial_{Q(x)}P(p-A(x))=0$ is the origin of the principle of least action. 

If there are some possible classical paths between space points 1 and 2, the quantum phases difference $e^{is}$ at 1 and 2 will be the summation of diffractive contribution by all possible classical paths [4, 3, 13]. In conclusion, it seems that the accurate equation of classical mechanics determines the quantum results. This is another expression of profound and intrinsic connection between classical mechanics and quantum theory.

Furthermore, because $\Omega=P(p-A(x))Q(x)$, any periodical symmetry condition of space $Q(x)$ would introduce to the invariance of phase, or $\Delta\Omega=2n\pi$. This fact is the origin of the quantization of mechanical values, such as $m, \epsilon, p, L, \nabla\times A$, etc. With the multi-order $n$ of high-spin fields, there will appear any kind of topological phase factors and flux quantization. This quantization rule, due to coordinate periodical symmetry condition of space-time, has already been found by Bohr et al, although its origin was not clear at that time [15]. 

\vspace{8pt}
\noindent {\Large \bf 8 Conclusion}
\vspace{4pt}

As demonstrated in the previous sections, the essential characteristic of nature is clearly the spinor system of space-time-matter. When expressed in the spinor combination of spae-time-matter, or the spinor pattern (1, 28), the state (2, 6, 13-1) and motion equation (3, 7, 13-2) of fields are exhibited in a clearer, simpler and preciser way; the source of the action and the principle of least action are evidently exposed (31); while the classical mechanics and quantum theory are thereby profoundly and intrinsically connected (23,32). It seems that nature is actually the existence and performance of a topological construction of space-time-matter in the spinor pattern.

\vspace{16pt}
\noindent {\Large \bf References}
\vspace{4pt}

[1] A.O.Barut, \it Electrodynamics and Classical Theory of Fields and Particles, \rm Dover Pub. Inc., New York 1980.

[2] P.A.M.Dirac, \it The Principles of Quantum Mechanics, \rm 4th ed., Oxford, Hong Kong 1958.

[3] W.Dittrich, M.Reuter,\it Classical and Quantum Dynamics, from Classical Path to Path Integrals, \rm Springer-Verlag, Berlin 1994.

[4] R.P.Feynman, A.R.Hibbs, \it Quantum Mechanics and Path Integrals, \rm McGraw-Hill, Inc., New York 1965.

[5] W.Greiner, \it Relativistic Quantum Mechanics, \rm Springer-Verlag, Berlin 1998.

[6] H.Goldstein, \it Classical Mechanics, \rm Addison-Wesleg, London 1980.

[7] F.Growss,\it Relativistic Quantum Mechanics and Field Theory, \rm John Wiley and Sons, London 1993.

[8] E. G.Harris, \it Introduction to Modern Theoretical Physics, \rm John Wiley and Sons, New York 1975.

[9] E. G.Harris, \it A Pedestrian Approach to Quantum Field Theory, \rm Wiley-Interscience, New York 1972.

[10] C.Itzykson, J. B.Zuber, \it Quantum Field Theory, \rm McGraw-Hill, New York 1980.

[11] J. D.Jackson, \it Classical Electrodynamics, \rm Wiley and Sons, New York 1976.

[12] B.Kaufman, `Crystal Statistics: Partition Function Evaluated by Spinor Analysis', \it Physical Review, \rm \bfseries76(8), \rm 1232(1949).

[13] H.Kleinert, \it Path Integrals in Quantum Mechanics, Statistics and Polymer Physics, \rm 2nd ed., World Scientific, Hong Kong 1995.

[14] L.D.Landau,  E.M.Lifshitz, \it Mechanics, \rm 3rd ed., Pergaman Press, New York 1976.

[15] E.Merzbacher, \it Quantum Mechanics, \rm 2nd ed., John Wiley and Sons, New York 1970.

[16] L.Ryder, \it Quantum Field Theory, \rm 2nd ed., Cambridge Uni. World-Scientific, New York 1996.

[17] S.Weinberg, \it The Quantum Theory of Fields, \rm Cambridge Uni, New York 1995.

[18] R. M.Wald, \it General Relativity, \rm The University of Chicago Press, Chicago and London 1984.

\end{document}